\documentclass[11pt, a4paper]{article}
\usepackage[utf8]{inputenc}
\usepackage{amsmath, setspace}
\usepackage{amsthm}
\usepackage{amsfonts}
\usepackage{mathtools}
\mathtoolsset{showonlyrefs}
\usepackage[shortlabels]{enumitem}
\usepackage{rotating}
\usepackage{pdflscape}
\usepackage{graphicx}
\usepackage{bbm}
\usepackage[dvipsnames]{xcolor}
\usepackage[colorlinks=true, linkcolor= RawSienna, citecolor = RawSienna, filecolor = RawSienna, urlcolor = RawSienna, hypertexnames = true, backref = page]{hyperref}
\usepackage[]{natbib} 
\bibpunct[:]{(}{)}{,}{a}{}{,}
\usepackage[english]{babel}
\usepackage{float}
\usepackage{caption}
\usepackage{subcaption}
\usepackage{tikz}
\usepackage{booktabs}
\usepackage{pdfpages}
\usepackage{threeparttable}
\usepackage{framed}
\usepackage{comment}
\usepackage{lscape}
\usepackage{bm}
\usepackage{adjustbox}
\usepackage[T1]{fontenc}
\usepackage{mlmodern}  
\DeclareFontFamily{OMX}{mlmex}{}
\DeclareFontShape{OMX}{mlmex}{m}{n}{<->mlmex10}{} 
\usepackage{tgtermes} 

\newtheorem{theorem}{Theorem}

\newtheorem{proposition}{Proposition}

\theoremstyle{remark}

\usepackage[margin=1in]{geometry}
\usepackage{sectsty}
\sectionfont{\centering}

\title{Conduct Parameter Estimation in Homogeneous Goods Markets with Equilibrium Existence and Uniqueness Conditions: The Case of Log-linear Specification}
\author{Yuri Matsumura\thanks{Department of Economics, Rice University. Email: \href{mailto:}{yuri.matsumura23@gmail.com}} \and Suguru Otani \thanks{Market Design Center, Department of Economics, University of Tokyo. Email: \href{mailto:}{suguru.otani@e.u-tokyo.ac.jp}.
\\Declarations of interest: none 
}}

\begin{document}

\maketitle
\begin{abstract}
    We propose a constrained generalized method of moments (GMM) estimator with some equilibrium uniqueness conditions for estimating the conduct parameter in a log-linear model with homogeneous goods markets.
    Monte Carlo simulations demonstrate that merely imposing parameter restrictions leads to not just inaccurate estimations but also some numerical issues, and adding the equilibrium uniqueness conditions resolves them.
    We also suggest a formulation of the GMM estimation to further avoid the numerical issues.
\end{abstract}

\vspace{0in}

\bigskip

\newpage

\section{Introduction}

Measuring competitiveness is central in empirical IO, and the conduct parameter is a widely used proxy.
Since marginal cost is typically unobserved, researchers identify and estimate the conduct parameter indirectly.

\citet{bresnahan1982oligopoly} develops an identification strategy under linear demand and marginal cost, and \citet{matsumura2023resolving} provide detailed conditions for this case.
Yet, empirical work often employs log-linear models \citep{okazaki2022excess,merel2009measuring}, which tend to produce implausibly low or negative estimates, even though identification conditions for general models are available from \citet{lau1982identifying} and \cite{matsumura2025revisiting}.
This casts doubt on the methodology and complicates model selection.

We address this by proposing a constrained GMM estimator that incorporates theoretical conditions for the uniqueness of equilibrium.
First, we derive new conditions guaranteeing a unique equilibrium.
Second, Monte Carlo simulations show that parameter restrictions alone yield inaccurate estimates and numerical errors, while adding equilibrium conditions resolves them.
We also propose a modified GMM formulation that further mitigates these issues.

\section{Model}
Consider data with $T$ markets with homogeneous products.
Assume there are $N$ firms in each market.
Let $t = 1,\ldots, T$ be the index of markets.
Then, we obtain the supply equation:
\begin{align}
     P_t + \theta Q_{t} P_t'(Q_{t})= MC_t(Q_{t}),\label{eq:supply_equation}
\end{align}
where $Q_{t}$ is the aggregate quantity, $P_t(Q_{t})$ is the inverse demand function, $MC_{t}(Q_{t})$ is the marginal cost function, and $\theta\in[0,1]$ is the conduct parameter. 
The equation nests perfect competition $(\theta=0)$, Cournot competition $(\theta=1 / N)$, and perfect collusion $(\theta=$ $1)$. See \citet{bresnahan1982oligopoly} for the details.

Consider an econometric model.
Assume that the inverse demand and the marginal cost functions are given as
\begin{align*}
    P_t = f(Q_{t}, X^{d}_{t}, \varepsilon^{d}_{t}, \alpha), 
    \\
    MC_t = g(Q_{t}, X^{c}_{t}, \varepsilon^{c}_{t}, \gamma),
\end{align*}
where $X^{d}_{t}$ and $X^{c}_{t}$ are the vector of exogenous variables, $\varepsilon^{d}_{t}$ and $\varepsilon^{c}_{t}$ are the error terms, and $\alpha$ and $\gamma$ are the vector of parameters.
We allow $X^{d}_{t}$ and $X^{c}_{t}$ to have common variables, but assume that there is at least one demand variable and one cost variable that are mutually excluded.
We also have the demand- and supply-side instruments, $Z^{d}_{t}$ and $Z^{c}_{t}$, and assume that the error terms satisfy the mean independence condition, $E[\varepsilon^{d}_{t}\mid X^{d}_{t}, Z^{d}_{t}] = E[\varepsilon^{c}_{t} \mid X^{c}_{t}, Z^{c}_{t}] =0$. 

The identification of the conduct parameter is indirectly characterized by \citet{lau1982identifying}:
\begin{theorem}\label{thm:lau_identification}
    Under the assumption that the industry inverse demand and cost functions are twice continuously differentiable, the index of competitiveness $\theta$ cannot be identified from data on industry price and output and other exogenous variables alone if and only if the industry inverse demand function is separable in $X^{d}$, that is, $f(Q, r(X^{d}))$, but not take the form $P = Q^{-1/\theta}r(X^{d}) + s(Q)$.
\end{theorem}
This theorem implies that the conduct parameter is identified if the inverse demand function is not separable.
A demand rotation instrument \citep{bresnahan1982oligopoly} achieves this.
See Appendix \ref{appendix:summary_goldman_uzawa} for the details of the definition of separability.

\subsection{Log-linear demand and log-linear marginal cost}
Consider a log-linear model, which is a typical specification.
The inverse demand and marginal cost functions are specified as
\begin{align}
    \log P_{t} &= \alpha_0 - (\alpha_1 + \alpha_2 Z^{R}_{t}) \log Q_t + \alpha_3 \log Y_t + \varepsilon^{d}_{t},\label{eq:log_linear_demand}\\
    \log MC_t &= \gamma_0 + \gamma_1 \log Q_t +  \gamma_2 \log W_{t} + \gamma_3 \log R_t + \varepsilon^{c}_{t},\label{eq:log_linear_marginal_cost}
\end{align}
where $Y_{t}$ and $Z_t^R$ are excluded demand shifters and $W_t$ and $R_t$ are excluded cost shifters.
When $Y_{t}$ and $Z_{t}^{R}$ vary without changing the equilibrium quantity, they work as the demand rotation instrument.  
Then, \eqref{eq:supply_equation} is written as
\begin{align}
    P_t &= \theta (\alpha_1 + \alpha_2 Z^{R}_{t}) P_t + MC_t.\label{eq:log_linear_supply_equation_direct}
\end{align}
By taking logarithm of \eqref{eq:log_linear_supply_equation_direct} and substituting \eqref{eq:log_linear_marginal_cost}, we obtain
\begin{align}
    \log P_t = - \log(1 - \theta(\alpha_1 + \alpha_2 Z^{R}_{t})) + \gamma_0 + \gamma_1 \log Q_t +  \gamma_2 \log W_{t} + \gamma_3 \log R_t + \varepsilon^{c}_{t}. \label{eq:log_linear_supply_equation}
\end{align}

The intersection of \eqref{eq:log_linear_demand} and \eqref{eq:log_linear_supply_equation} determines the equilibrium, but there could be multiple equilibria.
Although this model is widely known, no paper has examined the multiple equilibria problem to our knowledge.
The next proposition provides the conditions for uniqueness.
The proof is in the online appendix \ref{sec:appendix_proof}.
\begin{proposition}\label{prop:equilibrium_existence}
    Assume that $\alpha_1 + \alpha_2 Z^{R}\ne 0$. Let $\Xi = \gamma_0 + \gamma_1\frac{\alpha_0 + \alpha_3 \log Y + \varepsilon^{d}}{\alpha_1 + \alpha_2 Z^{R}} +  \gamma_2 \log W + \gamma_3 \log R + \varepsilon^{c}$.
    The number of equilibria is determined as follows:
    \begin{itemize}
        \item When $1 - \theta (\alpha_1 + \alpha_2 Z^{R}) \le 0$, there is no equilibrium,
        \item When $1 - \theta (\alpha_1 + \alpha_2 Z^{R}) >0$, 
        \begin{itemize}
            \item If $ \gamma_1 +\alpha_1+\alpha_2 Z^R \ne 0$, there is a unique equilibrium,
            \item If $\gamma_1 + \alpha_1+\alpha_2 Z^R = 0$, there are infinitely many equilibria when $\exp(\Xi) = 1 - \theta (\alpha_1 + \alpha_2 Z^{R}_{t})$, but there is no equilibrium otherwise.
        \end{itemize}
    \end{itemize}
\end{proposition}
The condition $1-\theta(\alpha_1+\alpha_2 Z^R)>0$ rules out the region where the log transformation leaves its domain, which corresponds to implausibly elastic demand combined with large conduct. The assumption $\gamma_1+\alpha_1+\alpha_2 Z^R\ne0$ excludes the knife-edge case in which the (pseudo) supply and demand are exactly parallel so that every price could be an equilibrium; both restrictions are technical and do not bind in regular empirical settings.

\section{Estimation}
Let $\xi = (\alpha_0,\alpha_1, \alpha_2, \alpha_3, \gamma_0,\gamma_1, \gamma_2, \gamma_3, \theta)$ be the vector of the parameters in the model. 
We use the GMM for the estimation.
Among GMM estimators, we apply the nonlinear system two-stage-least-squares (N2SLS) using \eqref{eq:log_linear_demand} and \eqref{eq:log_linear_supply_equation}.
We rewrite the demand equation \eqref{eq:log_linear_demand} and the supply equation \eqref{eq:log_linear_supply_equation} as
\begin{align}
    {\varepsilon}_t^d(\xi) & =  \log P_{t} - \alpha_0 + (\alpha_1 + \alpha_2 Z^{R}_{t}) \log Q_t - \alpha_3 \log Y_t \label{eq:residual_demand_2sls}, \\
    {\varepsilon}_t^c(\xi) & =  \log P_t + \log(1 - \theta(\alpha_1 + \alpha_2 Z^{R}_{t})) -\gamma_0 - \gamma_1 \log Q_t -  \gamma_2 \log W_{t} -\gamma_3 \log R_t \label{eq:residual_supply_2sls}.
\end{align}
To estimate the parameters, we convert the conditional moments, $E[\varepsilon_t^d\mid Z_t^d] = E[\varepsilon_t^c\mid Z_t^c]=0$, into unconditional moments, $E[\varepsilon_t^d Z_t^d] = E[\varepsilon_t^cZ_t^c]=0$.
Using Equations \eqref{eq:residual_demand_2sls} and \eqref{eq:residual_supply_2sls}, we construct the sample analog of the unconditional moments:
\begin{align*}
    g(\xi) = \left[\begin{array}{l}
    \frac{1}{T}\sum_{t=1}^T{\varepsilon}^{d}_{t}(\xi)Z_{t}^{d} \\
    \frac{1}{T}\sum_{t=1}^T{\varepsilon}^{c}_{t}(\xi)Z_{t}^{c}
    \end{array}\right].
\end{align*}
We define the N2SLS estimator as the solution to the problem,
\begin{align}
     \xi^* = \arg \min_{\xi}\ g(\xi)^\top W g(\xi) \label{eq:minimization_gmm}
\end{align}
where the weight matrix $W$ is defined as
\begin{align}
    W = \left[\frac{1}{T}\sum_{t = 1}^T Z_t^\top Z_t\right]^{-1} \text{ where } Z_{t}=\left[\begin{array}{ll}
        Z_{t}^{d\top} & 0 \\
        0 & Z_{t}^{c\top}
    \end{array}\right].\label{eq:weight_matrix}
\end{align}

We also add the following constraints based on Proposition \ref{prop:equilibrium_existence} to \eqref{eq:minimization_gmm}:
\begin{align}
    &0\le\theta \le 1,\label{eq:conduct_constraint}\\
    &\alpha_1 + \alpha_2 Z_{t}^{R} >0, \quad \gamma_1>0 ,\quad t = 1,\ldots, T\label{eq:slope_constraint}\\
    &1- \theta(\alpha_1 + \alpha_2 Z_{t}^{R}) >0,\quad t = 1,\ldots, T.\label{eq:equilibrium_existence}
\end{align}
Constraint \eqref{eq:conduct_constraint} is a standard assumption on the conduct parameter.
Constraint \eqref{eq:slope_constraint} implies the downward-sloping demand and upward-sloping marginal cost, which guarantees that $\gamma_1 + \alpha_1 + \alpha_2 Z^{R} \ne 0$.
Constraint \eqref{eq:equilibrium_existence} relates to the uniqueness of equilibrium. 
See the detailed simulation setting in the online appendix \ref{sec:setting}.

\section{Simulation results}\label{sec:results}

We compare N2SLS estimations with and without constraints in Table \ref{tb:loglinear_loglinear_sigma_1_simultaneous_non_constraint_theta_constraint_bias_rmse}.
Panel (a) shows that, without constraints, the estimator fails to recover $\gamma_0$ and $\theta$, replicating known issues due to the flat objective function and invalid search regions without equilibrium (Appendix \ref{appendix:implausible_estimator}).
Panel (b), which imposes Constraint \eqref{eq:conduct_constraint}, improves estimation in large samples via the domain restriction.\footnote{Constraints \eqref{eq:slope_constraint} and \eqref{eq:equilibrium_existence} alone yield severe bias; see Table \ref{tb:loglinear_loglinear_sigma_0.5_simultaneous_no_constraint_slope_constraint_bias_rmse} and Appendix \ref{sec:additional_experiments}.}
However, in small samples, demand parameter estimates degrade and convergence declines. When convergence fails, $\alpha_1$ becomes large, rendering $1 - \theta(\alpha_1 + \alpha_2 Z^R_t) < 0$ and causing numerical errors inside the log term in \eqref{eq:supply_equation}.
Adding constraints \eqref{eq:slope_constraint} and \eqref{eq:equilibrium_existence} in Panel (c) improves small-sample convergence and demand accuracy, though convergence is not guaranteed. 
In large samples, performance surpasses that of Panel (b) for some parameters.

To address convergence failure, we propose an alternative formulation (Table \ref{tb:loglinear_loglinear_sigma_1_mpec_theta_constraint_slope_constraint_bias_rmse}) that computes $\varepsilon^c_t$ via \eqref{eq:log_linear_marginal_cost} and enforces Equation \eqref{eq:log_linear_supply_equation_direct} as a constraint, along with Constraints \eqref{eq:conduct_constraint}–\eqref{eq:equilibrium_existence}.\footnote{See Appendix \ref{sec:setting} for details.}
This avoids log terms in both objective and constraints, achieving 100\% convergence and reducing $\theta$'s bias and RMSE to 0.014 and 0.217, though not dominating Panel (c) across all parameters.
In sum, incorporating equilibrium uniqueness conditions and eliminating log terms greatly improves conduct parameter estimation. Additional experiments appear in Appendix \ref{sec:additional_experiments}.

\begin{table}[!htbp]
  \begin{center}
  \caption{Performance comparison}
    \label{tb:loglinear_loglinear_sigma_1_simultaneous_non_constraint_theta_constraint_bias_rmse} 
  \text{(a) N2SLS without Constraints \eqref{eq:conduct_constraint}, \eqref{eq:slope_constraint}, and \eqref{eq:equilibrium_existence}}\\[0.5em]
  \begin{adjustbox}{width=\textwidth}
    
\begin{tabular}[t]{lrrrrrrrr}
\toprule
  & Bias & RMSE & Bias & RMSE & Bias & RMSE & Bias & RMSE\\
\midrule
$\alpha_{0}$ & -1.070 & 7.012 & -0.021 & 5.110 & 0.365 & 2.207 & 0.400 & 2.030\\
$\alpha_{1}$ & -0.164 & 1.060 & -0.001 & 0.782 & 0.073 & 0.458 & 0.096 & 0.574\\
$\alpha_{2}$ & -0.011 & 0.104 & -0.006 & 0.071 & 0.002 & 0.033 & 0.005 & 0.043\\
$\alpha_{3}$ & -0.101 & 0.619 & -0.005 & 0.474 & 0.021 & 0.198 & 0.029 & 0.187\\
$\gamma_{0}$ & 9.735 & 15.743 & 9.636 & 10.870 & 13.173 & 13.269 & 13.294 & 13.351\\
$\gamma_{1}$ & -0.070 & 1.624 & -0.177 & 0.469 & -0.184 & 0.248 & -0.177 & 0.220\\
$\gamma_{2}$ & -0.034 & 0.939 & -0.098 & 0.317 & -0.090 & 0.152 & -0.080 & 0.127\\
$\gamma_{3}$ & -0.047 & 0.750 & -0.091 & 0.311 & -0.098 & 0.156 & -0.085 & 0.133\\
$\theta$ & -3e+05 & 3e+06 & -2e+05 & 2e+06 & -8e+04 & 9e+04 & -9e+04 & 1e+05\\
Runs converged (\%) &  & 99.500 &  & 99.800 &  & 98.600 &  & 98.400\\
Sample size ($T$) &  & 100 &  & 200 &  & 1000 &  & 1500\\
\bottomrule
\end{tabular}

  \end{adjustbox}

  \vspace{1em}

  \text{(b) N2SLS with Constraints \eqref{eq:conduct_constraint}}\\[0.5em]
  \begin{adjustbox}{width=\textwidth}
    
\begin{tabular}[t]{lrrrrrrrr}
\toprule
  & Bias & RMSE & Bias & RMSE & Bias & RMSE & Bias & RMSE\\
\midrule
$\alpha_{0}$ & -1.922 & 8.603 & -0.068 & 5.116 & 0.037 & 2.035 & 0.000 & 1.556\\
$\alpha_{1}$ & -0.299 & 1.314 & -0.010 & 0.785 & 0.005 & 0.312 & 0.000 & 0.240\\
$\alpha_{2}$ & -0.013 & 0.104 & -0.002 & 0.063 & 0.001 & 0.024 & 0.000 & 0.019\\
$\alpha_{3}$ & -0.165 & 0.774 & -0.007 & 0.472 & -0.004 & 0.185 & -0.001 & 0.152\\
$\gamma_{0}$ & -1.767 & 14.394 & -1.001 & 6.530 & -0.208 & 1.993 & -0.156 & 1.566\\
$\gamma_{1}$ & 0.255 & 1.949 & 0.132 & 0.838 & 0.034 & 0.229 & 0.027 & 0.174\\
$\gamma_{2}$ & 0.125 & 1.097 & 0.053 & 0.475 & 0.017 & 0.150 & 0.019 & 0.119\\
$\gamma_{3}$ & 0.099 & 0.903 & 0.062 & 0.481 & 0.007 & 0.149 & 0.014 & 0.120\\
$\theta$ & -0.098 & 0.441 & -0.060 & 0.421 & -0.061 & 0.319 & -0.058 & 0.281\\
Runs converged (\%) &  & 98.100 &  & 98.700 &  & 100.000 &  & 100.000\\
Sample size ($T$) &  & 100 &  & 200 &  & 1000 &  & 1500\\
\bottomrule
\end{tabular}

  \end{adjustbox}

  \vspace{1em}

  \text{(c) N2SLS with Constraints \eqref{eq:conduct_constraint}, \eqref{eq:slope_constraint}, and \eqref{eq:equilibrium_existence}}\\[0.5em]
  \begin{adjustbox}{width=\textwidth}
    
\begin{tabular}[t]{lrrrrrrrr}
\toprule
  & Bias & RMSE & Bias & RMSE & Bias & RMSE & Bias & RMSE\\
\midrule
$\alpha_{0}$ & -0.905 & 6.954 & 0.120 & 5.001 & 0.072 & 2.042 & 0.052 & 1.563\\
$\alpha_{1}$ & -0.141 & 1.053 & 0.018 & 0.768 & 0.010 & 0.313 & 0.008 & 0.241\\
$\alpha_{2}$ & -0.006 & 0.101 & 0.000 & 0.062 & 0.001 & 0.024 & 0.001 & 0.019\\
$\alpha_{3}$ & -0.088 & 0.620 & 0.007 & 0.475 & -0.001 & 0.186 & 0.003 & 0.152\\
$\gamma_{0}$ & -1.748 & 14.206 & -0.938 & 6.428 & 0.015 & 1.995 & 0.163 & 1.570\\
$\gamma_{1}$ & 0.254 & 1.927 & 0.129 & 0.825 & 0.018 & 0.226 & 0.003 & 0.170\\
$\gamma_{2}$ & 0.117 & 1.083 & 0.049 & 0.467 & 0.008 & 0.148 & 0.007 & 0.116\\
$\gamma_{3}$ & 0.098 & 0.890 & 0.058 & 0.478 & -0.001 & 0.148 & 0.003 & 0.118\\
$\theta$ & -0.100 & 0.441 & -0.072 & 0.424 & -0.121 & 0.351 & -0.148 & 0.333\\
Runs converged (\%) &  & 99.600 &  & 99.900 &  & 100.000 &  & 100.000\\
Sample size ($T$) &  & 100 &  & 200 &  & 1000 &  & 1500\\
\bottomrule
\end{tabular}

  \end{adjustbox}
  \end{center}
  \footnotesize
  Note: The error terms are drawn from a normal distribution, $N(0, \sigma)$. True values: $\alpha_0=20.0, \alpha_1=1.0, \alpha_2=0.1, \alpha_3=1.0, \gamma_0=5.0, \gamma_1=1.0, \gamma_2=1.0, \gamma_3=1.0, \theta=0.5$ and $\sigma=1.0$. See online appendix \ref{sec:setting} and \cite{matsumura2024challenges} for the setting.
\end{table}

\begin{table}[!htbp]
  \begin{center}
  \caption{Ad hoc method using \eqref{eq:log_linear_marginal_cost} to compute $\varepsilon_t^c$ and \eqref{eq:log_linear_supply_equation_direct} with Constraints \eqref{eq:conduct_constraint}, \eqref{eq:slope_constraint}, and \eqref{eq:equilibrium_existence}}
  \label{tb:loglinear_loglinear_sigma_1_mpec_theta_constraint_slope_constraint_bias_rmse} 
  \begin{adjustbox}{width=\textwidth}
    
\begin{tabular}[t]{lrrrrrrrr}
\toprule
  & Bias & RMSE & Bias & RMSE & Bias & RMSE & Bias & RMSE\\
\midrule
$\alpha_{0}$ & -0.614 & 5.995 & -0.213 & 4.315 & 0.077 & 2.034 & 0.063 & 1.555\\
$\alpha_{1}$ & -0.085 & 0.902 & -0.024 & 0.663 & 0.011 & 0.312 & 0.010 & 0.240\\
$\alpha_{2}$ & -0.028 & 0.105 & -0.022 & 0.073 & 0.000 & 0.025 & 0.001 & 0.020\\
$\alpha_{3}$ & -0.070 & 0.549 & -0.019 & 0.431 & -0.001 & 0.185 & 0.004 & 0.152\\
$\gamma_{0}$ & -5.106 & 15.922 & -2.379 & 6.990 & -0.375 & 1.959 & -0.398 & 1.533\\
$\gamma_{1}$ & 0.386 & 2.047 & 0.141 & 0.839 & 0.045 & 0.229 & 0.044 & 0.175\\
$\gamma_{2}$ & 0.190 & 1.155 & 0.054 & 0.475 & 0.022 & 0.150 & 0.027 & 0.120\\
$\gamma_{3}$ & 0.163 & 1.006 & 0.065 & 0.482 & 0.013 & 0.149 & 0.023 & 0.121\\
$\theta$ & 0.186 & 0.442 & 0.158 & 0.422 & -0.007 & 0.275 & 0.014 & 0.217\\
Runs converged (\%) &  & 100.000 &  & 100.000 &  & 100.000 &  & 100.000\\
Sample size ($T$) &  & 100 &  & 200 &  & 1000 &  & 1500\\
\bottomrule
\end{tabular}

  \end{adjustbox}
  \end{center}
\end{table}

\section{Discussion}
Two concerns surround the conduct parameter approach: the difficulty of interpreting intermediate or extreme values, and the critique by \citet{corts1999conduct} that it may understate market power under collusion.\footnote{As \citet{magnolfi2022comparison} note, this critique does not apply when the data stem from a static model.}
We show that implausible estimates in log-linear models often stem from numerical issues—especially when equilibrium conditions are omitted—rather than conceptual flaws. 
Addressing these issues yields more stable and interpretable results.

This distinction matters: misattributing numerical artifacts to theoretical limits risks dismissing a useful tool. Our findings aim to encourage more constructive use of the conduct parameter approach.

\paragraph{Acknowledgments}
We thank Jeremy Fox, Yelda Gungor, and Isabelle Perrigne for their valuable comments. 
We were supported by JST ERATO Grant Number JPMJER2301 and Otani was supported by JSPS Grant-in-Aid (KAKENHI) for Young Researcher JSPS 24K22604 and 25K16620.

\newpage
\bibliographystyle{aer}
\bibliography{conduct_parameter}

\newpage
\appendix

\section{Online appendix}

\subsection{Summary of Goldman and Uzawa (1964)}\label{appendix:summary_goldman_uzawa}

\citet{goldmanNote1964} investigate separability concepts in demand analysis.
Let $N$ be a set of variables, $N = \{1,\ldots, n\}$.
Consider a partition of $N$ into $S$ parts, $\{N^1, \ldots, N^S\}$ such that $N = \bigcup_{s=1}^S N^s$ and $N^s \cap N^t = \emptyset$.

A function is weakly separable with respect to a partition if 
\begin{align}
    \frac{\partial}{\partial x_k}\left(\frac{\partial f(x)/\partial x_i}{\partial f(x)/\partial x_j} \right)= 0, \quad i,j\in N^s, k \notin N^s.
\end{align}
This implies that, for example, the marginal rate of substitution between $i$ and $j$ in the same partition is independent of the quantities of commodities outside $N^s$.

In Theorem 2, \citet{goldmanNote1964} characterizes the function that satisfies the weak separability.
\begin{theorem}[Theorem 2 in \citet{goldmanNote1964}]
    A function $f(x)$ is weakly separable with respect to a partition $\{N^1, .. ., N^s\}$ if, and only if, $f(x)$ is of the form: 
    \begin{align}
        f(X) = \Phi(f^1(x^{(1)}),\ldots, f^s(x^{(s)})   )
    \end{align} where $\Phi(f^1,\ldots, f^s)$ is a function of $S$ variables and, for each $s$, $f^s(x^{(s)})$ is a function subvector $x^{(s)}$ alone.
\end{theorem}

Let $\{Q, X_{1}^{d},\ldots, X_{K}^{d}\}$ be the set of dependent variables and divide the variables into two groups $(Q, X^{d})$.
Then, as the inverse demand function $P(Q, X^{d}) = f(Q, r(X^{d}))$ in Theorem \ref{thm:lau_identification} has the form in the above theorem, we can say that the conduct parameter is not identified when the inverse demand function is separable.

In contrast, we can see that the linear demand function in \citet{bresnahan1982oligopoly} and the log-linear demand function in this paper are not separable.
For the linear case,
\begin{align}
    \frac{\partial P(Q, X^{d})/\partial Z^{R}}{\partial P(Q, X^{d})/\partial Y} = \frac{ \alpha_2 Q}{\alpha_3} \Longrightarrow \frac{\partial}{\partial Q}\left(\frac{\partial P(Q, X^{d})/\partial Z^{R}}{\partial P(Q, X^{d})/\partial Y}\right) = \frac{\alpha_2}{\alpha_3} \ne 0.
\end{align}
The last part, $\frac{\alpha_2}{\alpha_3} \ne 0$, is concluded from the identification condition in \citet{matsumura2023resolving}.
For the log-linear case, we have
\begin{align}
    \frac{\partial P(Q, X^{d})/\partial Z^{R}}{\partial P(Q, X^{d})/\partial Y}  = \frac{\alpha_2 Y \log Q}{\alpha_3} \Longrightarrow \frac{\partial}{\partial Q}\left(\frac{\partial P(Q, X^{d})/\partial Z^{R}}{\partial P(Q, X^{d})/\partial Y}\right) = \frac{\alpha_2Y}{\alpha_3Q} \ne 0.
\end{align}
Thus, again when $\frac{\alpha_2}{\alpha_3}$ is nonzero, which emphasizes the role of the demand shifters,  the inverse demand function is not separable.
Note that due to the log specification, $Y$ and $Q$ are positive.

\subsection{Existence and uniqueness of equilibrium prices}\label{sec:appendix_proof}

Proposition \ref{prop:equilibrium_existence} proposes the conditions for the unique existence of $P_{t}(>0)$ solving the demand equation \eqref{eq:log_linear_demand} and supply equation \eqref{eq:log_linear_supply_equation} for $P_{t}$ under $\theta\in[0,1]$.
The proof is not based on the optimization by individual firms but checks if there exists a point at which the demand function and the marginal cost function cross.
Therefore, we allow an equilibrium price to exist when the demand and the marginal cost are upward-sloping.
    
\begin{proof}
    
Rewriting the demand equation \eqref{eq:log_linear_demand} as 
\begin{align*}
    \log Q_{t}(P_{t})= \frac{\alpha_0 - \log P_{t} + \alpha_3 \log Y_t + \varepsilon^{d}_{t}}{(\alpha_1 + \alpha_2 Z^{R}_{t})}   
\end{align*}
and substituting this into the supply equation \eqref{eq:log_linear_supply_equation}, we obtain
\begin{align}
    P_t &=\theta (\alpha_1 + \alpha_2 Z^{R}_{t}) P_t + \exp\left(\gamma_0 + \gamma_1 \log Q_t(P_{t}) +  \gamma_2 \log W_{t} + \gamma_3 \log R_t + \varepsilon^{c}_{t}\right). \nonumber\\
    & = \theta(\alpha_1 + \alpha_2 Z^{R}_{t})P_t + \exp\left(\gamma_0 + \gamma_1 \frac{\alpha_0 - \log P_{t} + \alpha_3 \log Y_t + \varepsilon^{d}_{t}}{(\alpha_1 + \alpha_2 Z^{R}_{t})} +\gamma_2 \log W_{t} + \gamma_3 \log R_{t} + \varepsilon^{c}_{t} \right)\nonumber\\
    & = \theta(\alpha_1 + \alpha_2 Z^{R}_{t})P_t  + \exp\left(\Xi + \frac{-\gamma_1}{\alpha_1+\alpha_2 Z^{R}_t} \log P_t \right)\nonumber\\
    &= \theta(\alpha_1 + \alpha_2 Z^{R}_{t})P_t  + \exp(\Xi) P_t^{\frac{-\gamma_1}{\alpha_1 + \alpha_2 Z^{R}_{t}}} \label{eq:equilibrium_equation}
\end{align}
where $\Xi = \gamma_0 + \gamma_1\frac{\alpha_0 + \alpha_3 \log Y_t + \varepsilon^{d}_{t}}{\alpha_1 + \alpha_2 Z^{R}_{t}} +  \gamma_2 \log W_{t} + \gamma_3 \log R_t + \varepsilon^{c}_{t}$.

Any price that satisfies \eqref{eq:equilibrium_equation} becomes an equilibrium price.
To find an equilibrium price $P_{t}^*$, we define $\Delta(P_t)$ as follows:
\begin{align}
    \Delta(P_t)
    &= \underbrace{[1 - \theta (\alpha_1 + \alpha_2 Z^{R}_{t})]P_t}_{(I)} - \underbrace{\exp(\Xi) P_t^{\frac{-\gamma_1}{\alpha_1 + \alpha_2 Z^{R}_{t}}}}_{(II)} \label{eq:fixed_point}.
\end{align}

We label the first term as (I) and the second term as (II).

An important note is that $P_t =0$ satisfies $\Delta(P_{t}) = 0$.
Thus $P_t^* = 0$ holds.
However, our interest is a unique positive equilibrium price, so we seek conditions in which there is a unique positive price that satisfies $\Delta(P_t)=0$.
    
When $1 - \theta (\alpha_1 + \alpha_2 Z^{R}_{t}) \le 0$, (I) is always negative on $P_t >0$. 
In contrast, (II) is non-negative regardless of the sign of $-\gamma_1/(\alpha_1+\alpha_2 Z^R)$ on $P_t > 0$.
Therefore, $\Delta(P_t)$ is always negative on $P_t>0$, which implies that there is no positive equilibrium price.

When $1 - \theta (\alpha_1 + \alpha_2 Z^{R}_{t}) \le 0$, (I) becomes a line passing through the origin with a positive slope, illustrated in Figure \ref{fg:equilibrium_existence}. 
Note that the first and second derivatives of (II) are given as
\begin{align}
    \frac{d}{dP_t}\exp(\Xi) P_t^{\frac{-\gamma_1}{\alpha_1 + \alpha_2 Z^{R}_{t}}} &= -\frac{\gamma_1}{\alpha_1 + \alpha_2 Z^R_t} \exp(\Xi)P_t^{-\frac{\gamma_1}{\alpha_1 + \alpha_2 Z^R_t} - 1},\\
    \frac{d^2}{dP_t^2} \exp(\Xi) P_t^{\frac{-\gamma_1}{\alpha_1 + \alpha_2 Z^{R}_{t}}} &= -\frac{\gamma_1}{\alpha_1 + \alpha_2 Z^R_t}\left(-\frac{\gamma_1}{\alpha_1 + \alpha_2 Z^R_t} - 1\right) \exp(\Xi) P_t^{-\frac{\gamma_1}{\alpha_1 + \alpha_2 Z^R_t} - 2}.
\end{align}
Therefore, the shape of (II) on $P_t >0$ is changed based on the value of $-\gamma_1/(\alpha_1+\alpha_2 Z^R)$ and the number that (I) and (II) cross on $P_t >0$ is also determined.

The case $-\gamma_1/(\alpha_1+\alpha_2 Z^R) < 0$ is illustrated in Figure \ref{fig:fixed_monotone_decreasing}. In this case, (II) is a monotone decreasing convex function because the first and second derivatives are both negative. As (I) is monotone increasing in $P_t >0$, (I) and (II) cross only once on $P_t >0$.

The case $-\gamma_1/(\alpha_1+\alpha_2 Z^R) \in [0, 1)$ is illustrated in Figure \ref{fig:fixed_concave}. In this case, (II) becomes a monotone increasing concave function passing through the origin because the first derivative is positive and the second derivative is negative.
As the value of the first derivative is infinity at $P_t = 0$ since $-\gamma_1/(\alpha_1+\alpha_2 Z^R)-1<0$, (II) should be greater than (I) around $P_t = 0$. 
Then, as $P_t$ becomes large starting from zero, the difference in both terms becomes small, and both terms cross only once on $P_t >0$ eventually.

When $-\gamma_1/(\alpha_1+\alpha_2 Z^R) = 1$, \eqref{eq:fixed_point} becomes
\[ 0 = [1 - \theta (\alpha_1 + \alpha_2 Z^{R}_{t}) - \exp(\Xi)] P_t. \]
If $ 1 - \theta (\alpha_1 + \alpha_2 Z^{R}_{t}) = \exp(\Xi)$, the above equation holds for any $P_t$. 
Thus, there are infinitely many equilibrium prices on $P_t >0$, which is illustrated in Figure \ref{fig:fixed_coinside}.
When $ 1 - \theta (\alpha_1 + \alpha_2 Z^{R}_{t}) \ne \exp(\Xi)$, there is no equilibrium price on $P_t >0$ because the right-hand side of the equation is always non-zero on $P_t >0$, which is illustrated in Figure \ref{fig:fixed_parallel}.

The case $-\gamma_1/(\alpha_1+\alpha_2 Z^R) > 1$ is illustrated in Figure \ref{fig:fixed_convex}. In this case, (II) becomes a monotone increasing convex function passing through the origin because the first and second derivatives are positive. Therefore, (I) and (II) cross only once on $P_t >0$.

\end{proof}

\begin{figure}[!ht]
    \caption{The illustrations of the equilibrium existence and uniqueness conditions}
    \label{fg:equilibrium_existence}
    \begin{subfigure}{0.45\textwidth}
        \begin{tikzpicture}
            \draw[->,thick] (-0.5,0)--(4,0) node[right]{$P_t$};
            \draw[->,thick] (0,-0.5)--(0,4) ;
            \draw[dotted, thick] (0.1, 3.9) .. controls (1, 1.5)..(4,0.5) node[above]{\scriptsize{$\exp(\Xi) P_t^{\frac{-\gamma_1}{\alpha_1 + \alpha_2 Z^{R}_{t}}}$}};
            \draw[thick] (0,0) -- (4,3) node[above]{\scriptsize{$[1-\theta(\alpha_1+\alpha_2 Z_t^R)]P_t$}};
        \end{tikzpicture}
        \caption{$\frac{-\gamma_1}{\alpha_1 + \alpha_2 Z^{R}_{t}} < 0$}
        \label{fig:fixed_monotone_decreasing}
    \end{subfigure}
    \hfill
    \begin{subfigure}{0.45\textwidth}
        \begin{tikzpicture}
            \draw[->,thick] (-0.5,0)--(4,0) node[right]{$P_t$};
            \draw[->,thick] (0,-0.5)--(0,4) ;
            \draw[dotted, thick] (0,0) .. controls (1, 2.0)..(4,2.5);
            \draw[thick] (0,0) -- (4,3);
        \end{tikzpicture}
         \caption{$\frac{-\gamma_1}{\alpha_1 + \alpha_2 Z^{R}_{t}} \in [0,1)$}
         \label{fig:fixed_concave}
    \end{subfigure}
    
    \vspace{1em}
    \begin{subfigure}{0.45\textwidth}
        \begin{tikzpicture}
            \draw[->,thick] (-0.5,0)--(4,0) node[right]{$P_t$};
            \draw[->,thick] (0,-0.5)--(0,4) ;
            \draw[dotted, thick] (0,0) -- (4,3);
            \draw[thin] (0,0) -- (4,3);
        \end{tikzpicture}
         \caption{$\frac{-\gamma_1}{\alpha_1 + \alpha_2 Z^{R}_{t}} = 1, 1-\theta(\alpha_1+\alpha_2 Z_t^R) = \exp(\Xi) $}
         \label{fig:fixed_coinside}
    \end{subfigure}
    \hfill
    \begin{subfigure}{0.45\textwidth}
        \begin{tikzpicture}
            \draw[->,thick] (-0.5,0)--(4,0) node[right]{$P_t$};
            \draw[->,thick] (0,-0.5)--(0,4) ;
            \draw[dotted, thick] (0,0) -- (4,3.9);
            \draw[thin] (0,0) -- (4,3);
        \end{tikzpicture}
         \caption{$\frac{-\gamma_1}{\alpha_1 + \alpha_2 Z^{R}_{t}} = 1, 1-\theta(\alpha_1+\alpha_2 Z_t^R) \ne \exp(\Xi) $}
         \label{fig:fixed_parallel}
    \end{subfigure}
    \vspace{1em}
    \begin{subfigure}{0.4\textwidth}
        \begin{tikzpicture}
            \draw[->,thick] (-0.5,0)--(4,0) node[right]{$P_t$};
            \draw[->,thick] (0,-0.5)--(0,4) ;
            \draw[dotted, thick] (0,0) .. controls (2, 0.5)..(4,3.5);
            \draw[thin] (0,0) -- (4,3);
        \end{tikzpicture}
         \caption{$\frac{-\gamma_1}{\alpha_1 + \alpha_2 Z^{R}_{t}}> 1$}
         \label{fig:fixed_convex}
    \end{subfigure}
    
    \footnotesize
    Note: As there is no equilibrium when $1- \theta(\alpha_1 + \alpha_2 Z^{R}_{t}) \le 0$, the above figures assume that $1- \theta(\alpha_1 + \alpha_2 Z^{R}_{t}) > 0$.
    The thick line represents the first term and the dotted line represents the second term in \eqref{eq:fixed_point}.
\end{figure}

From $\Delta (P_t) = 0$, the positive equilibrium price can be written as 
\begin{align}
    0 & = [1-\theta(\alpha_1 + \alpha_2 Z^{R}_{t})]P_t - \exp(\Xi) P_t^{\frac{-\gamma_1}{\alpha_1 + \alpha_2 Z^{R}_{t}}}\nonumber \\ 
    0 & = 1-\theta(\alpha_1 + \alpha_2 Z^{R}_{t}) - \exp(\Xi)P_t^{\frac{-\gamma_1}{\alpha_1 + \alpha_2 Z^{R}_{t}}- 1} \nonumber\\ 
    P_t^{\frac{-\gamma_1}{\alpha_1 + \alpha_2 Z^{R}_{t}}- 1} & = \frac{1-\theta(\alpha_1 + \alpha_2 Z^{R}_{t})}{ \exp(\Xi)}\nonumber\\ 
    P_{t}^* &= \left(\frac{1-\theta(\alpha_1 + \alpha_2 Z^{R}_{t})}{\exp(\Xi)}\right)^{-\frac{\alpha_1 + \alpha_2 Z^{R}_{t}}{\gamma_1 +\alpha_1 + \alpha_2 Z^{R}_{t}}}.
\end{align} 

Figure \ref{fg:equilibrium_existence_demand_supply} illustrates how the demand and supply equations cross under the conditions.
When $1- \theta(\alpha_1 + \alpha_2 Z^{R}_{t}) \le 0$, the supply equation \eqref{eq:log_linear_supply_equation} is ill-defined because the inside of the log function becomes negative.
Thus, there should not be any equilibrium.
Hereafter, assume that $1- \theta(\alpha_1 + \alpha_2 Z^{R}_{t}) > 0$.
When $-\gamma_1/(\alpha_1 + \alpha_2 Z^{R}_{t}) = 1$ and $\exp(\Xi) = 1- \theta(\alpha_1 + \alpha_2 Z^{R}_{t})$, the demand equation and supply equation coincides.
Thus there are infinitely many equilibria in the model.
When $-\gamma_1/(\alpha_1 + \alpha_2 Z^{R}_{t}) = 1$ and $\exp(\Xi) \ne 1- \theta(\alpha_1 + \alpha_2 Z^{R}_{t})$, the demand and supply equations have a same slope and different intercepts, which means that both equations become parallel.
Thus, there is no equilibrium.
When $-\gamma_1/(\alpha_1 + \alpha_2 Z^{R}_{t}) \ne 1$, the demand and supply equations have different slopes, and hence we can find a unique equilibrium.

\begin{figure}[!ht]
    \caption{The illustrations of the equilibrium existence and uniqueness conditions}
    \label{fg:equilibrium_existence_demand_supply}
    \begin{subfigure}{0.45\textwidth}
        \begin{tikzpicture}
            \draw[->,thick] (-0.5,0)--(4,0) node[right]{$Q_t$};
            \draw[->,thick] (0,-0.5)--(0,4) node[above]{$P_t$};
            \draw[blue, thick] (0,0.5) .. controls (1, 2.2)..(4,3) node[right]{$S$};
            \draw[red, thin] (0,0.5) .. controls (1, 2.2)..(4,3) node[above]{$D$};
        \end{tikzpicture}
         \caption{$-\gamma_1/(\alpha_1 + \alpha_2 Z^{R}_{t}) = 1$ and $\exp(\Xi) = 1- \theta(\alpha_1 + \alpha_2 Z^{R}_{t})$}
    \end{subfigure}
    \hfill
    \begin{subfigure}{0.45\textwidth}
        \begin{tikzpicture}
            \draw[->,thick] (-0.5,0)--(4,0) node[right]{$Q_t$};
            \draw[->,thick] (0,-0.5)--(0,4) node[above]{$P_t$};
            \draw[blue, thick] (0, 0.5) .. controls (1, 2.2)..(4,3) node[right]{$S$};
            \draw[red, thick, yshift = 5mm] (0, 0.5) .. controls (1, 2.2)..(4,3) node[right]{$D$};
        \end{tikzpicture}
        \caption{$-\gamma_1/(\alpha_1 + \alpha_2 Z^{R}_{t}) = 1$ and $\exp(\Xi) \ne 1- \theta(\alpha_1 + \alpha_2 Z^{R}_{t})$}
    \end{subfigure}
    
    \vspace{1em}
    \begin{subfigure}{0.45\textwidth}
        \begin{tikzpicture}
            \draw[->,thick] (-0.5,0)--(4,0) node[right]{$Q_t$};
            \draw[->,thick] (0,-0.5)--(0,4) node[above]{$P_t$};
            \draw[blue, thick] (0,0.5) .. controls (1, 2.2)..(4,3) node[right]{$S$};
            \draw[red, thick] (0,3.8) .. controls (1, 1.5) .. (4,1.0)  node[right]{$D$};
        \end{tikzpicture}
         \caption{$\gamma_1/(\alpha_1 + \alpha_2 Z^{R}_{t}) \ne 1$, $\alpha_1 + \alpha_2 Z^{R}_{t}>0$, and $\gamma_1>0$}
    \end{subfigure}
    \hfill
    \begin{subfigure}{0.45\textwidth}
        \begin{tikzpicture}
            \draw[->,thick] (-0.5,0)--(4,0) node[right]{$Q_t$};
            \draw[->,thick] (0,-0.5)--(0,4) node[above]{$P_t$};
            \draw[blue, thick] (0,0.5) .. controls (1, 2.2)..(4,3.2) node[right]{$S$};
            \draw[red, thick] (0,1.4) .. controls (1, 2.2) .. (4,2.8)  node[below]{$D$};
        \end{tikzpicture}
         \caption{$\gamma_1/(\alpha_1 + \alpha_2 Z^{R}_{t}) \ne 1$, $\alpha_1 + \alpha_2 Z^{R}_{t}\le0$, and $\gamma_1>0$}
    \end{subfigure}
    \vspace{1em}
    \begin{subfigure}{0.45\textwidth}
        \begin{tikzpicture}
            \draw[->,thick] (-0.5,0)--(4,0) node[right]{$Q_t$};
            \draw[->,thick] (0,-0.5)--(0,4) node[above]{$P_t$};
            \draw[blue, thick] (0,3.0) .. controls (1, 2.0)..(4,1.5) node[right]{$S$};
            \draw[red, thick] (0,3.8) .. controls (1, 1.5) .. (4,1.0)  node[right]{$D$};
        \end{tikzpicture}
         \caption{$\gamma_1/(\alpha_1 + \alpha_2 Z^{R}_{t}) \ne 1$, $\alpha_1 + \alpha_2 Z^{R}_{t}>0$, and $\gamma_1\le 0$}
    \end{subfigure}
    \hfill
    \begin{subfigure}{0.45\textwidth}
        \begin{tikzpicture}
            \draw[->,thick] (-0.5,0)--(4,0) node[right]{$Q_t$};
            \draw[->,thick] (0,-0.5)--(0,4) node[above]{$P_t$};
            \draw[red, thick] (0,0.5) .. controls (1, 2.2)..(4,3) node[right]{$D$};
            \draw[blue, thick] (0,3.8) .. controls (1, 1.5) .. (4,1.0)  node[right]{$S$};
        \end{tikzpicture}
         \caption{$\gamma_1/(\alpha_1 + \alpha_2 Z^{R}_{t}) \ne 1$, $\alpha_1 + \alpha_2 Z^{R}_{t}\le 0$, and $\gamma_1\le 0$}
    \end{subfigure}
    
    \footnotesize
    Note: As there is no equilibrium when $1- \theta(\alpha_1 + \alpha_2 Z^{R}_{t}) \le 0$, the above figures assume that $1- \theta(\alpha_1 + \alpha_2 Z^{R}_{t}) > 0$. 
    Case (a) and (b) also allow both the demand and supply equations to be downward sloping.
\end{figure}

\newpage

\subsection{Simulation and estimation procedure}\label{sec:setting}
To generate the simulation data, for each model, we first generate the exogenous variables $Y_t, Z^{R}_{t}, W_t, R_{t}, H_t$, and $K_t$ and the error terms $\varepsilon_{t}^c$ and $\varepsilon_{t}^d$ based on the data generation process in Table \ref{tb:parameter_setting}.
By substituting the Equation \eqref{eq:log_linear_demand} into Equation \eqref{eq:log_linear_supply_equation} and solving it for $P_{t}$, the log aggregate quantity is given as: 
\begin{align}
    \log Q_t &= \frac{ \alpha_0 + \alpha_3 \log Y_t + \log (1 - \theta (\alpha_1 + \alpha_2 Z^{R}_{t})) - \gamma_0  -  \gamma_2 \log W_{t} - \gamma_3 \log R_t + \varepsilon^{d}_{t} - \varepsilon^{c}_{t}}{\gamma_1+ \alpha_1 + \alpha_2 Z^{R}_{t} }.\label{eq:quantity_loglinear}
\end{align}
We compute the equilibrium quantity $Q_{t}$ for the log-linear model by \eqref{eq:quantity_loglinear}.
We then compute the equilibrium price $P_t$ by substituting $Q_{t}$ and other variables into the demand function \eqref{eq:log_linear_demand}.
We generate 1000 data sets of 100, 200, 1000, 1500 markets.
We jointly estimate the demand and supply parameters by the simultaneous equation model \citep{wooldridge2010econometric} from the true values.
We use state-of-the-art constrained optimization solvers, i.e., \texttt{Ipopt.jl} which implements an interior point line search filter method that aims to find a local solution of nonlinear programming problems.

\begin{table}[!htbp]
    \caption{True parameters and distributions}
    \label{tb:parameter_setting}
    \begin{center}
    \subfloat[Parameters]{
    \begin{tabular}{crr}
            \hline
            $\alpha_0$  & $20.0$ &\\
            $\alpha_1$ & $1.0$  &\\
            $\alpha_2$ & $0.1$ &\\
            $\alpha_3$ & $1.0$ &\\
            $\gamma_0$ & $5.0$  &\\
            $\gamma_1$ & $1.0$  &\\
            $\gamma_2$ & $1.0$ &\\
            $\gamma_3$ & $1.0$ &\\
            $\theta$ & $0.5$  &\\
            \hline
        \end{tabular}
    }
    \subfloat[Distributions]{
    \begin{tabular}{crr}
            \hline
            Demand shifter&  &  \\
            $Y_t$ & $N(0,1)$ \\
            Demand rotation instrument&  &  \\
            $Z^{R}_{t}$ & $U(0,1)$ \\
            Cost shifter  &  \\
            $W_{t}$ & $U(1,3)$ \\
            $R_t$  & $U(1,3)$  \\
            $H_{t}$ & $W_{t}+U(0,1)$  \\
            $K_{t}$ & $R_{t}+U(0,1)$  \\
            Error&  &  \\
            $\varepsilon^{d}_{t}$ & $N(0,\sigma)$  \\
            $\varepsilon^{c}_{t}$ & $N(0,\sigma)$ \\
            \hline
        \end{tabular}
    }
    \end{center}
    \footnotesize
    Note: $\sigma=\{0.5, 1.0, 2.0\}$. $N:$ Normal distribution. $U:$ Uniform distribution.
\end{table}

The ad-hoc estimation for Table \ref{tb:loglinear_loglinear_sigma_1_mpec_theta_constraint_slope_constraint_bias_rmse} is formulated as follows.
Let $\Phi = (\xi, \{MC_t\}_{t = 1,\ldots, T})$.
Consider a constrained minimization problem such that 
\begin{align}
         &\Phi^{*} = \arg\min_{\Phi}\ g(\Phi)^\top W g(\Phi),\\
\end{align}
subject to
\begin{align}
     &(\Phi) = \left[\begin{array}{l}
    \frac{1}{T}\sum_{t=1}^T{\varepsilon}^{d}_{t}(\Phi)Z_{t}^{d} \\
    \frac{1}{T}\sum_{t=1}^T{\varepsilon}^{c}_{t}(\Phi)Z_{t}^{c}
    \end{array}\right],\\ 
    & {\varepsilon}_t^d(\Phi) =  \log P_{t} - \alpha_0 + (\alpha_1 + \alpha_2 Z^{R}_{t}) \log Q_t - \alpha_3 \log Y_t, \\
    & {\varepsilon}_t^c(\Phi) =  \log MC_t -\gamma_0 - \gamma_1 \log Q_t -  \gamma_2 \log W_{t} -\gamma_3 \log R_t, \\
    & P_{t} = \theta(\alpha_1 + \alpha_2 Z^{R}_t)P_{t} + MC_{t},\\
    & MC_t \ge 0,
\end{align}
where the matrix $W$ is the same as the original N2SLS.

\subsection{Why does the estimation without equilibrium conditions converge to the extremely low conduct parameters?}\label{appendix:implausible_estimator}

In the literature, it is often observed that estimation without imposing equilibrium conditions tends to converge to negative values of the conduct parameter.
We confirm this pattern in our Monte Carlo simulations, particularly in Panel (a) of Table \ref{tb:loglinear_loglinear_sigma_1_simultaneous_non_constraint_theta_constraint_bias_rmse}.
To understand this phenomenon, consider the supply equation \eqref{eq:log_linear_supply_equation}, which includes the following term:
\begin{align}
-\log(1 - \theta (\alpha_1 + \alpha_2 Z^{R}{t})) + \gamma{0}.
\end{align}
If the model allows for $\theta < 0$, the term inside the logarithm can become arbitrarily large in magnitude by choosing a sufficiently negative $\theta$ when $\alpha_1 + \alpha_2 Z^{R}_{t} > 0$:
\begin{align}
\lim_{\theta \rightarrow -\infty} \log(1 - \theta (\alpha_1 + \alpha_2 Z^{R}_{t})) \rightarrow \infty.
\end{align}
However, since $\theta$ appears inside a logarithmic function, changes in $\theta$—even when large—have a relatively small impact on the GMM objective function. Moreover, the effect of changes in $\theta$ can be offset by adjusting $\gamma_0$, with only modest changes in its value.
This implies that the GMM objective function can be very flat in the neighborhood of the true values of $\theta$ and $\gamma_0$.

\begin{figure}
    \begin{center}
    \includegraphics[width=0.7\linewidth]{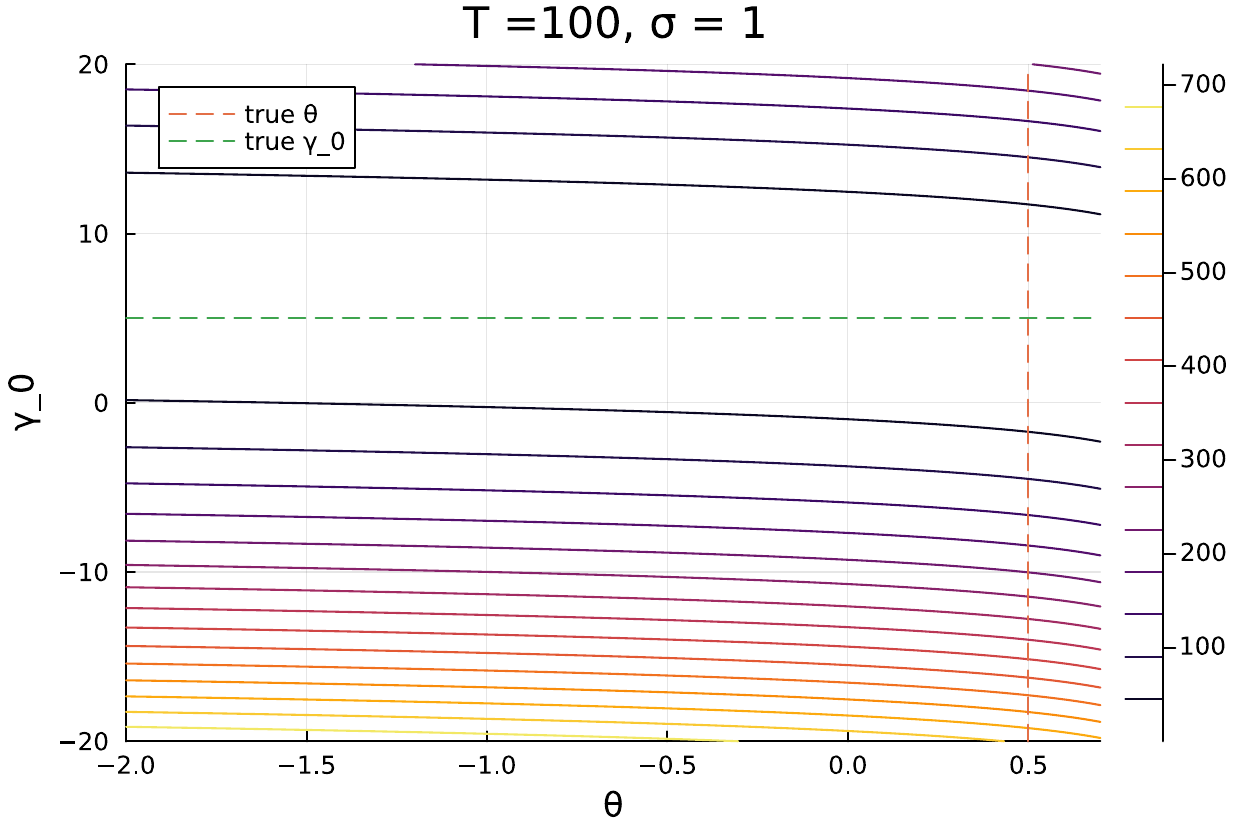}
    \caption{The contour map of the GMM objective function on $\theta - \gamma_0$ plane.}
    \label{fig:gmm_contour}    
    \end{center}
    \footnotesize
    Note: We pick one simulation data where $T = 100$ and $\sigma  =1$. The GMM objective function is flat around the true values of the intercept of the log marginal cost and the conduct parameter.
\end{figure}

We verify this implication by examining the shape of the GMM objective function.
Figure \ref{fig:gmm_contour} plots the objective function on the $\theta$–$\gamma_0$ plane, fixing all other parameters at their true values and using one simulated dataset.
The same pattern is observed across other simulation datasets as well.
As the figure clearly shows, the GMM objective function is nearly flat around the true value of $\gamma_0$, and it allows for a wide range of $\theta$ values, especially in the negative direction, without significantly affecting the objective function value.

\subsection{Additional experiments}\label{sec:additional_experiments}

Additional results for different $\sigma$ are shown in Tables \ref{tb:loglinear_loglinear_sigma_0.5_simultaneous_theta_constraint_slope_constraint_bias_rmse} and \ref{tb:loglinear_loglinear_sigma_2_simultaneous_theta_constraint_non_constraint_bias_rmse}. 
As a summary, the main findings in the main text are robust.
 
Table \ref{tb:loglinear_loglinear_sigma_2_mpec_theta_constraint_non_constraint_bias_rmse} shows the results of the estimation with ado hoc method with different standard deviations of the error terms.
As in Table \ref{tb:loglinear_loglinear_sigma_1_mpec_theta_constraint_slope_constraint_bias_rmse}, the run convergence rate becomes 100\% and the accuracy of the conduct parameter estimation is improved.
However, again, the results with ad hoc improvement do not dominate the results without the improvement for all parameters.

In Tables \ref{tb:linear_linear_sigma_1} and  \ref{tb:linear_linear_sigma_2_mpec_linear_non_constraint_theta_constraint_bias_rmse}, we illustrate that N2SLS with Constraints \eqref{eq:conduct_constraint} works for the linear model as in \cite{matsumura2023resolving}. 
This means that incorporating the conditions is innocuous for the linear model. 

Table \ref{tb:loglinear_loglinear_sigma_0.5_simultaneous_no_constraint_slope_constraint_bias_rmse} demonstrates that relying solely on Constraints \eqref{eq:slope_constraint} and \eqref{eq:equilibrium_existence} within the N2SLS framework leads to severe bias, implying that only Constraints \eqref{eq:slope_constraint} and \eqref{eq:equilibrium_existence} cannot resolve the problem. Table \ref{tb:loglinear_loglinear_sigma_0.5_simultaneous_theta_constraint_non_constraint_bias_rmse} demonstrates that relying solely on Constraints \eqref{eq:conduct_constraint} within the N2SLS framework leads to severe bias—especially in small samples—for parameters other than the conduct parameter, as well as poor convergence. While the conduct parameter itself appears stable, this is mechanically due to the imposed domain constraint, which prevents extreme estimates by construction. These results underscore that Constraints \eqref{eq:conduct_constraint} alone are insufficient and highlight the critical importance of incorporating the equilibrium conditions \eqref{eq:slope_constraint} and \eqref{eq:equilibrium_existence}. In large samples, however, the problem is substantially alleviated, as the parameter search becomes less problematic.

\begin{table}[!htbp]
  \centering
  \caption{Performance comparison ($\sigma=0.5$)}
  \label{tb:loglinear_loglinear_sigma_0.5_simultaneous_theta_constraint_slope_constraint_bias_rmse}
  \text{(a)N2SLS without Constraints \eqref{eq:conduct_constraint}, \eqref{eq:slope_constraint}, and \eqref{eq:equilibrium_existence}}\\[0.5em]
  \begin{adjustbox}{width=0.95\textwidth}
    
\begin{tabular}[t]{lrrrrrrrr}
\toprule
  & Bias & RMSE & Bias & RMSE & Bias & RMSE & Bias & RMSE\\
\midrule
$\alpha_{0}$ & 0.119 & 4.055 & 0.346 & 3.180 & 0.344 & 1.497 & 0.379 & 1.582\\
$\alpha_{1}$ & 0.019 & 0.620 & 0.054 & 0.486 & 0.076 & 0.429 & 0.099 & 0.576\\
$\alpha_{2}$ & 0.000 & 0.050 & 0.003 & 0.039 & 0.002 & 0.029 & 0.003 & 0.036\\
$\alpha_{3}$ & 0.009 & 0.336 & 0.025 & 0.268 & 0.028 & 0.129 & 0.031 & 0.129\\
$\gamma_{0}$ & 10.104 & 11.066 & 9.397 & 10.288 & 13.240 & 13.280 & 13.314 & 13.346\\
$\gamma_{1}$ & -0.146 & 0.340 & -0.166 & 0.264 & -0.180 & 0.198 & -0.181 & 0.192\\
$\gamma_{2}$ & -0.072 & 0.229 & -0.079 & 0.165 & -0.087 & 0.108 & -0.086 & 0.099\\
$\gamma_{3}$ & -0.069 & 0.230 & -0.082 & 0.171 & -0.088 & 0.110 & -0.087 & 0.100\\
$\theta$ & -5e+04 & 1e+05 & -3e+04 & 6e+04 & -8e+04 & 9e+04 & -8e+04 & 9e+04\\
Runs converged (\%) &  & 100.000 &  & 100.000 &  & 98.100 &  & 98.700\\
Sample size ($T$) &  & 100 &  & 200 &  & 1000 &  & 1500\\
\bottomrule
\end{tabular}

  \end{adjustbox}
  \vspace{1.5em}  
  \text{(b) N2SLS with Constraints \eqref{eq:conduct_constraint}, \eqref{eq:slope_constraint}, and \eqref{eq:equilibrium_existence}}\\[0.5em]
  \begin{adjustbox}{width=0.95\textwidth}
   
\begin{tabular}[t]{lrrrrrrrr}
\toprule
  & Bias & RMSE & Bias & RMSE & Bias & RMSE & Bias & RMSE\\
\midrule
$\alpha_{0}$ & -0.006 & 3.556 & 0.175 & 3.158 & 0.059 & 0.977 & -0.017 & 0.763\\
$\alpha_{1}$ & -0.002 & 0.545 & 0.026 & 0.484 & 0.009 & 0.150 & -0.003 & 0.117\\
$\alpha_{2}$ & 0.002 & 0.044 & 0.004 & 0.037 & 0.000 & 0.012 & 0.000 & 0.010\\
$\alpha_{3}$ & 0.001 & 0.309 & 0.012 & 0.266 & 0.005 & 0.091 & 0.000 & 0.073\\
$\gamma_{0}$ & -0.616 & 3.317 & -0.221 & 2.191 & 0.275 & 1.076 & 0.421 & 0.972\\
$\gamma_{1}$ & 0.089 & 0.399 & 0.038 & 0.252 & -0.015 & 0.107 & -0.027 & 0.089\\
$\gamma_{2}$ & 0.044 & 0.265 & 0.022 & 0.170 & -0.007 & 0.074 & -0.011 & 0.061\\
$\gamma_{3}$ & 0.046 & 0.273 & 0.019 & 0.178 & -0.008 & 0.078 & -0.012 & 0.059\\
$\theta$ & -0.064 & 0.390 & -0.079 & 0.339 & -0.119 & 0.260 & -0.145 & 0.261\\
Runs converged (\%) &  & 99.900 &  & 100.000 &  & 100.000 &  & 100.000\\
Sample size ($T$) &  & 100 &  & 200 &  & 1000 &  & 1500\\
\bottomrule
\end{tabular}

  \end{adjustbox}
  \footnotesize
\end{table}

\begin{table}[!htbp]
  \centering
  \caption{Performance comparison ($\sigma=2.0$)}
  \label{tb:loglinear_loglinear_sigma_2_simultaneous_theta_constraint_non_constraint_bias_rmse}

  \text{(a) N2SLS without Constraints \eqref{eq:conduct_constraint}, \eqref{eq:slope_constraint}, and \eqref{eq:equilibrium_existence}}\\[0.5em]
  \begin{adjustbox}{width=0.95\textwidth}
    
\begin{tabular}[t]{lrrrrrrrr}
\toprule
  & Bias & RMSE & Bias & RMSE & Bias & RMSE & Bias & RMSE\\
\midrule
$\alpha_{0}$ & -2.358 & 11.286 & -1.457 & 8.696 & 0.231 & 4.747 & 0.450 & 4.028\\
$\alpha_{1}$ & -0.360 & 1.770 & -0.227 & 1.313 & 0.040 & 0.744 & 0.095 & 0.711\\
$\alpha_{2}$ & -0.021 & 0.138 & -0.012 & 0.142 & 0.002 & 0.063 & 0.006 & 0.065\\
$\alpha_{3}$ & -0.175 & 0.967 & -0.138 & 0.867 & 0.006 & 0.434 & 0.031 & 0.347\\
$\gamma_{0}$ & 9.868 & 21.289 & 10.490 & 13.359 & 13.151 & 13.453 & 13.177 & 13.384\\
$\gamma_{1}$ & -0.015 & 2.542 & -0.130 & 1.042 & -0.202 & 0.400 & -0.186 & 0.339\\
$\gamma_{2}$ & -0.057 & 1.656 & -0.083 & 0.735 & -0.090 & 0.274 & -0.099 & 0.224\\
$\gamma_{3}$ & -0.020 & 2.115 & -0.065 & 0.719 & -0.097 & 0.279 & -0.092 & 0.232\\
$\theta$ & -7e+05 & 5e+06 & -3e+05 & 2e+06 & -1e+05 & 4e+05 & -9e+04 & 1e+05\\
Runs converged (\%) &  & 99.600 &  & 99.800 &  & 98.300 &  & 98.900\\
Sample size ($T$) &  & 100 &  & 200 &  & 1000 &  & 1500\\
\bottomrule
\end{tabular}

  \end{adjustbox}

  \vspace{1.5em}  

  \text{(b) N2SLS with Constraints \eqref{eq:conduct_constraint}, \eqref{eq:slope_constraint}, and \eqref{eq:equilibrium_existence}}\\[0.5em]
  \begin{adjustbox}{width=0.95\textwidth}
    
\begin{tabular}[t]{lrrrrrrrr}
\toprule
  & Bias & RMSE & Bias & RMSE & Bias & RMSE & Bias & RMSE\\
\midrule
$\alpha_{0}$ & -2.008 & 11.292 & -1.149 & 8.680 & 0.145 & 4.655 & 0.274 & 3.838\\
$\alpha_{1}$ & -0.308 & 1.773 & -0.181 & 1.313 & 0.020 & 0.713 & 0.041 & 0.590\\
$\alpha_{2}$ & -0.016 & 0.136 & -0.007 & 0.139 & 0.003 & 0.057 & 0.004 & 0.043\\
$\alpha_{3}$ & -0.147 & 0.985 & -0.113 & 0.875 & -0.003 & 0.429 & 0.017 & 0.333\\
$\gamma_{0}$ & 0.661 & 11.910 & -1.090 & 13.475 & -0.665 & 4.501 & -0.404 & 3.245\\
$\gamma_{1}$ & -0.044 & 1.600 & 0.172 & 1.770 & 0.098 & 0.556 & 0.066 & 0.391\\
$\gamma_{2}$ & -0.056 & 1.086 & 0.088 & 1.172 & 0.054 & 0.346 & 0.024 & 0.252\\
$\gamma_{3}$ & -0.020 & 1.205 & 0.108 & 1.225 & 0.049 & 0.365 & 0.031 & 0.267\\
$\theta$ & -0.229 & 0.475 & -0.163 & 0.464 & -0.090 & 0.410 & -0.108 & 0.394\\
Runs converged (\%) &  & 99.200 &  & 99.200 &  & 99.700 &  & 100.000\\
Sample size ($T$) &  & 100 &  & 200 &  & 1000 &  & 1500\\
\bottomrule
\end{tabular}

  \end{adjustbox}
  \footnotesize
\end{table}

\begin{table}[!htbp]
  \centering
  \caption{Ad hoc method using \eqref{eq:log_linear_marginal_cost} to compute $\varepsilon_t^c$ and \eqref{eq:log_linear_supply_equation_direct} as constraints with Constraints \eqref{eq:conduct_constraint}, \eqref{eq:slope_constraint}, and \eqref{eq:equilibrium_existence}}
  \label{tb:loglinear_loglinear_sigma_2_mpec_theta_constraint_non_constraint_bias_rmse}

  \text{(a) $\sigma = 0.5$}\\[0.5em]
  \begin{adjustbox}{width=0.95\textwidth}
    
\begin{tabular}[t]{lrrrrrrrr}
\toprule
  & Bias & RMSE & Bias & RMSE & Bias & RMSE & Bias & RMSE\\
\midrule
$\alpha_{0}$ & -0.347 & 2.978 & -0.153 & 2.547 & 0.065 & 0.976 & -0.009 & 0.760\\
$\alpha_{1}$ & -0.043 & 0.452 & -0.016 & 0.388 & 0.010 & 0.150 & -0.001 & 0.117\\
$\alpha_{2}$ & -0.024 & 0.065 & -0.017 & 0.054 & 0.000 & 0.012 & 0.000 & 0.010\\
$\alpha_{3}$ & -0.028 & 0.265 & -0.013 & 0.224 & 0.005 & 0.091 & 0.001 & 0.073\\
$\gamma_{0}$ & -2.458 & 5.660 & -1.258 & 2.959 & -0.132 & 0.942 & -0.153 & 0.748\\
$\gamma_{1}$ & 0.112 & 0.602 & 0.057 & 0.260 & 0.014 & 0.103 & 0.014 & 0.082\\
$\gamma_{2}$ & 0.053 & 0.314 & 0.030 & 0.172 & 0.007 & 0.073 & 0.010 & 0.060\\
$\gamma_{3}$ & 0.061 & 0.426 & 0.028 & 0.181 & 0.006 & 0.077 & 0.008 & 0.058\\
$\theta$ & 0.169 & 0.407 & 0.109 & 0.356 & -0.003 & 0.155 & 0.017 & 0.116\\
Runs converged (\%) &  & 100.000 &  & 100.000 &  & 100.000 &  & 100.000\\
Sample size ($T$) &  & 100 &  & 200 &  & 1000 &  & 1500\\
\bottomrule
\end{tabular}

  \end{adjustbox}

  \vspace{1em}
  \text{(c) $\sigma = 2.0$}\\[0.5em]
  \begin{adjustbox}{width=0.95\textwidth}
    
\begin{tabular}[t]{lrrrrrrrr}
\toprule
  & Bias & RMSE & Bias & RMSE & Bias & RMSE & Bias & RMSE\\
\midrule
$\alpha_{0}$ & -0.835 & 9.908 & -0.561 & 6.310 & 0.147 & 4.424 & 0.301 & 3.686\\
$\alpha_{1}$ & -0.109 & 1.563 & -0.076 & 0.935 & 0.024 & 0.677 & 0.047 & 0.566\\
$\alpha_{2}$ & -0.043 & 0.124 & -0.033 & 0.135 & -0.005 & 0.061 & 0.001 & 0.045\\
$\alpha_{3}$ & -0.047 & 0.909 & -0.078 & 0.664 & -0.004 & 0.404 & 0.018 & 0.323\\
$\gamma_{0}$ & -6.629 & 20.418 & -5.373 & 16.692 & -1.445 & 4.738 & -1.123 & 3.411\\
$\gamma_{1}$ & 0.422 & 2.630 & 0.384 & 2.103 & 0.137 & 0.580 & 0.115 & 0.409\\
$\gamma_{2}$ & 0.184 & 1.788 & 0.185 & 1.318 & 0.073 & 0.357 & 0.048 & 0.261\\
$\gamma_{3}$ & 0.199 & 1.885 & 0.215 & 1.496 & 0.068 & 0.376 & 0.055 & 0.276\\
$\theta$ & 0.225 & 0.459 & 0.227 & 0.447 & 0.098 & 0.357 & 0.084 & 0.313\\
Runs converged (\%) &  & 99.200 &  & 99.800 &  & 99.800 &  & 100.000\\
Sample size ($T$) &  & 100 &  & 200 &  & 1000 &  & 1500\\
\bottomrule
\end{tabular}

  \end{adjustbox}
  \footnotesize
\end{table}

\begin{table}[!htbp]
  \begin{center}
  \caption{N2SLS without Constraints \eqref{eq:conduct_constraint} results of the linear model}
  \label{tb:linear_linear_sigma_1} 

  \text{(a) $\sigma = 0.5$}\\[0.5em]
  \begin{adjustbox}{width=0.95\textwidth}
    
\begin{tabular}[t]{llrrrrrrr}
\toprule
  & Bias & RMSE & Bias & RMSE & Bias & RMSE & Bias & RMSE\\
\midrule
$\alpha_{0}$ & -0.018 & 0.465 & 0.007 & 0.323 & -0.008 & 0.213 & -0.006 & 0.097\\
$\alpha_{1}$ & -0.045 & 2.257 & 0.024 & 1.523 & 0.018 & 1.016 & -0.031 & 0.455\\
$\alpha_{2}$ & -0.001 & 0.255 & -0.001 & 0.176 & -0.004 & 0.115 & 0.001 & 0.051\\
$\alpha_{3}$ & -0.005 & 0.108 & 0.003 & 0.075 & -0.001 & 0.050 & -0.001 & 0.022\\
$\gamma_{0}$ & -0.061 & 0.732 & -0.005 & 0.474 & -0.021 & 0.346 & -0.005 & 0.152\\
$\gamma_{1}$ & -0.311 & 3.450 & -0.124 & 1.928 & -0.081 & 1.303 & -0.003 & 0.548\\
$\gamma_{2}$ & 0.009 & 0.109 & -0.001 & 0.071 & 0.003 & 0.051 & 0.000 & 0.023\\
$\gamma_{3}$ & 0.001 & 0.108 & 0.003 & 0.075 & 0.003 & 0.053 & 0.000 & 0.022\\
$\theta$ & 0.047 & 0.354 & 0.017 & 0.209 & 0.014 & 0.135 & 0.003 & 0.058\\
Sample size ($T$) &  & 50 &  & 100 &  & 200 &  & 1000\\
\bottomrule
\end{tabular}

  \end{adjustbox}

  \vspace{1em}

  \text{(b) $\sigma = 1.0$}\\[0.5em]
  \begin{adjustbox}{width=0.95\textwidth}
    
\begin{tabular}[t]{llrrrrrrr}
\toprule
  & Bias & RMSE & Bias & RMSE & Bias & RMSE & Bias & RMSE\\
\midrule
$\alpha_{0}$ & -0.027 & 1.023 & -0.002 & 0.641 & -0.004 & 0.448 & -0.016 & 0.188\\
$\alpha_{1}$ & -0.024 & 4.396 & -0.169 & 2.965 & 0.061 & 2.060 & 0.011 & 0.905\\
$\alpha_{2}$ & -0.006 & 0.494 & 0.016 & 0.325 & -0.007 & 0.234 & -0.006 & 0.100\\
$\alpha_{3}$ & -0.006 & 0.223 & 0.002 & 0.153 & -0.001 & 0.099 & -0.003 & 0.045\\
$\gamma_{0}$ & -0.318 & 1.769 & -0.091 & 1.059 & -0.086 & 0.714 & -0.004 & 0.308\\
$\gamma_{1}$ & 5.859 & 210.853 & -0.679 & 6.280 & -0.338 & 2.972 & -0.050 & 1.110\\
$\gamma_{2}$ & 0.035 & 0.247 & 0.011 & 0.157 & 0.011 & 0.104 & 0.000 & 0.045\\
$\gamma_{3}$ & 0.045 & 0.250 & 0.012 & 0.150 & 0.010 & 0.104 & 0.002 & 0.045\\
$\theta$ & -0.399 & 18.450 & 0.098 & 0.738 & 0.054 & 0.308 & 0.009 & 0.113\\
Sample size ($T$) &  & 50 &  & 100 &  & 200 &  & 1000\\
\bottomrule
\end{tabular}

  \end{adjustbox}

  \vspace{1em}

  \text{(c) $\sigma = 2.0$}\\[0.5em]
  \begin{adjustbox}{width=0.95\textwidth}
    
\begin{tabular}[t]{llrrrrrrr}
\toprule
  & Bias & RMSE & Bias & RMSE & Bias & RMSE & Bias & RMSE\\
\midrule
$\alpha_{0}$ & -0.263 & 2.596 & 0.071 & 1.670 & -0.040 & 0.947 & -0.002 & 0.412\\
$\alpha_{1}$ & -0.271 & 10.820 & 0.008 & 6.492 & 0.236 & 4.263 & 0.021 & 1.809\\
$\alpha_{2}$ & -0.044 & 1.253 & 0.023 & 0.779 & -0.031 & 0.483 & -0.003 & 0.210\\
$\alpha_{3}$ & -0.024 & 0.584 & 0.008 & 0.343 & -0.004 & 0.225 & 0.003 & 0.092\\
$\gamma_{0}$ & -2.074 & 19.624 & -0.551 & 3.043 & -0.171 & 1.516 & -0.051 & 0.633\\
$\gamma_{1}$ & 58.209 & 1750.688 & -2.416 & 56.909 & -3.617 & 39.044 & -0.103 & 2.334\\
$\gamma_{2}$ & 0.242 & 2.430 & 0.065 & 0.409 & 0.020 & 0.220 & 0.006 & 0.093\\
$\gamma_{3}$ & 0.230 & 2.328 & 0.055 & 0.404 & 0.010 & 0.219 & 0.008 & 0.092\\
$\theta$ & -6.668 & 233.851 & 0.372 & 6.334 & 0.418 & 3.820 & 0.024 & 0.245\\
Sample size ($T$) &  & 50 &  & 100 &  & 200 &  & 1000\\
\bottomrule
\end{tabular}

  \end{adjustbox}
        \end{center}
  \footnotesize
  Note: The data generating process follows \cite{matsumura2023resolving}.
\end{table}


\begin{table}[!htbp]
  \begin{center}
  \caption{N2SLS with Constraints \eqref{eq:conduct_constraint} for the linear model}
  \label{tb:linear_linear_sigma_2_mpec_linear_non_constraint_theta_constraint_bias_rmse} 

  \text{(a) $\sigma = 0.5$}\\[0.5em]
  \begin{adjustbox}{width=0.95\textwidth}
    
\begin{tabular}[t]{llrrrrrrr}
\toprule
  & Bias & RMSE & Bias & RMSE & Bias & RMSE & Bias & RMSE\\
\midrule
$\alpha_{0}$ & -0.013 & 0.462 & 0.008 & 0.322 & -0.008 & 0.213 & -0.006 & 0.097\\
$\alpha_{1}$ & -0.096 & 2.201 & 0.015 & 1.511 & 0.018 & 1.016 & -0.031 & 0.455\\
$\alpha_{2}$ & 0.006 & 0.247 & 0.001 & 0.174 & -0.004 & 0.115 & 0.001 & 0.051\\
$\alpha_{3}$ & -0.004 & 0.108 & 0.003 & 0.074 & -0.001 & 0.050 & -0.001 & 0.022\\
$\gamma_{0}$ & -0.054 & 0.724 & -0.002 & 0.472 & -0.021 & 0.346 & -0.005 & 0.152\\
$\gamma_{1}$ & -0.098 & 2.620 & -0.093 & 1.847 & -0.081 & 1.303 & -0.003 & 0.548\\
$\gamma_{2}$ & 0.008 & 0.108 & -0.002 & 0.070 & 0.003 & 0.051 & 0.000 & 0.023\\
$\gamma_{3}$ & 0.001 & 0.107 & 0.003 & 0.075 & 0.003 & 0.053 & 0.000 & 0.022\\
$\theta$ & 0.023 & 0.258 & 0.014 & 0.197 & 0.014 & 0.135 & 0.003 & 0.058\\
Sample size ($T$) &  & 50 &  & 100 &  & 200 &  & 1000\\
\bottomrule
\end{tabular}

  \end{adjustbox}

  \vspace{1em}

  \text{(b) $\sigma = 1.0$}\\[0.5em]
  \begin{adjustbox}{width=0.95\textwidth}
    
\begin{tabular}[t]{llrrrrrrr}
\toprule
  & Bias & RMSE & Bias & RMSE & Bias & RMSE & Bias & RMSE\\
\midrule
$\alpha_{0}$ & 0.012 & 1.015 & 0.012 & 0.636 & 0.001 & 0.446 & -0.016 & 0.188\\
$\alpha_{1}$ & -0.383 & 4.216 & -0.291 & 2.872 & 0.015 & 2.012 & 0.011 & 0.905\\
$\alpha_{2}$ & 0.042 & 0.466 & 0.033 & 0.310 & -0.001 & 0.227 & -0.006 & 0.100\\
$\alpha_{3}$ & 0.000 & 0.222 & 0.004 & 0.152 & 0.000 & 0.099 & -0.003 & 0.045\\
$\gamma_{0}$ & -0.233 & 1.612 & -0.067 & 1.020 & -0.076 & 0.703 & -0.004 & 0.308\\
$\gamma_{1}$ & 0.406 & 4.332 & -0.024 & 3.174 & -0.178 & 2.585 & -0.050 & 1.110\\
$\gamma_{2}$ & 0.025 & 0.231 & 0.008 & 0.154 & 0.010 & 0.103 & 0.000 & 0.045\\
$\gamma_{3}$ & 0.034 & 0.233 & 0.010 & 0.146 & 0.009 & 0.103 & 0.002 & 0.045\\
$\theta$ & 0.014 & 0.380 & 0.018 & 0.311 & 0.035 & 0.260 & 0.009 & 0.113\\
Sample size ($T$) &  & 50 &  & 100 &  & 200 &  & 1000\\
\bottomrule
\end{tabular}

  \end{adjustbox}

  \vspace{1em}

  \text{(c) $\sigma = 2.0$}\\[0.5em]
  \begin{adjustbox}{width=0.95\textwidth}
    
\begin{tabular}[t]{llrrrrrrr}
\toprule
  & Bias & RMSE & Bias & RMSE & Bias & RMSE & Bias & RMSE\\
\midrule
$\alpha_{0}$ & -0.138 & 2.592 & 0.142 & 1.671 & 0.004 & 0.936 & 0.000 & 0.410\\
$\alpha_{1}$ & -0.986 & 11.090 & -0.665 & 6.325 & -0.174 & 4.042 & 0.003 & 1.788\\
$\alpha_{2}$ & 0.065 & 1.255 & 0.112 & 0.756 & 0.023 & 0.447 & 0.000 & 0.207\\
$\alpha_{3}$ & -0.006 & 0.589 & 0.019 & 0.345 & 0.003 & 0.224 & 0.003 & 0.092\\
$\gamma_{0}$ & -0.123 & 3.350 & -0.298 & 2.455 & -0.081 & 1.379 & -0.047 & 0.628\\
$\gamma_{1}$ & 0.635 & 7.424 & 0.141 & 5.668 & -0.081 & 4.255 & -0.040 & 2.159\\
$\gamma_{2}$ & 0.016 & 0.484 & 0.037 & 0.349 & 0.009 & 0.205 & 0.005 & 0.093\\
$\gamma_{3}$ & 0.009 & 0.517 & 0.026 & 0.336 & 0.000 & 0.208 & 0.007 & 0.092\\
$\theta$ & -0.029 & 0.446 & 0.025 & 0.414 & 0.034 & 0.381 & 0.016 & 0.220\\
Sample size ($T$) &  & 50 &  & 100 &  & 200 &  & 1000\\
\bottomrule
\end{tabular}

  \end{adjustbox}
      
  \end{center}
  \footnotesize
  Note: The data generating process follows \cite{matsumura2023resolving}.
\end{table}


\begin{table}[!htbp]
  \centering
  \caption{N2SLS with Constraints \eqref{eq:slope_constraint} and \eqref{eq:equilibrium_existence} for the loglinear model}
  \label{tb:loglinear_loglinear_sigma_0.5_simultaneous_no_constraint_slope_constraint_bias_rmse} 

  \text{(a) $\sigma = 0.5$}\\[0.5em]
  \begin{adjustbox}{width=0.95\textwidth}
    
\begin{tabular}[t]{lrrrrrrrr}
\toprule
  & Bias & RMSE & Bias & RMSE & Bias & RMSE & Bias & RMSE\\
\midrule
$\alpha_{0}$ & 0.066 & 3.646 & 0.536 & 3.302 & 0.638 & 2.013 & 0.544 & 1.606\\
$\alpha_{1}$ & 0.011 & 0.559 & 0.084 & 0.508 & 0.122 & 0.451 & 0.103 & 0.383\\
$\alpha_{2}$ & 0.001 & 0.048 & 0.004 & 0.040 & 0.006 & 0.037 & 0.006 & 0.042\\
$\alpha_{3}$ & 0.006 & 0.317 & 0.039 & 0.280 & 0.049 & 0.161 & 0.044 & 0.131\\
$\gamma_{0}$ & 8.768 & 9.771 & 10.695 & 11.255 & 12.797 & 12.854 & 12.890 & 12.928\\
$\gamma_{1}$ & -0.143 & 0.364 & -0.173 & 0.261 & -0.182 & 0.203 & -0.180 & 0.192\\
$\gamma_{2}$ & -0.069 & 0.247 & -0.081 & 0.164 & -0.089 & 0.111 & -0.086 & 0.099\\
$\gamma_{3}$ & -0.067 & 0.239 & -0.085 & 0.170 & -0.089 & 0.111 & -0.087 & 0.100\\
$\theta$ & -19426.126 & 60643.314 & -34723.006 & 57457.978 & -64958.969 & 153375.492 & -61625.065 & 71909.012\\
Runs converged (\%) &  & 100.000 &  & 100.000 &  & 96.500 &  & 93.400\\
Sample size ($T$) &  & 100 &  & 200 &  & 1000 &  & 1500\\
\bottomrule
\end{tabular}

  \end{adjustbox}

  \vspace{1em}

  \text{(b) $\sigma = 1.0$}\\[0.5em]
  \begin{adjustbox}{width=0.95\textwidth}
    
\begin{tabular}[t]{lrrrrrrrr}
\toprule
  & Bias & RMSE & Bias & RMSE & Bias & RMSE & Bias & RMSE\\
\midrule
$\alpha_{0}$ & -1.036 & 7.004 & 0.110 & 5.112 & 0.576 & 3.164 & 0.582 & 2.252\\
$\alpha_{1}$ & -0.158 & 1.057 & 0.022 & 0.782 & 0.116 & 0.682 & 0.122 & 0.638\\
$\alpha_{2}$ & -0.009 & 0.103 & 0.000 & 0.102 & 0.007 & 0.050 & 0.008 & 0.056\\
$\alpha_{3}$ & -0.098 & 0.615 & 0.007 & 0.481 & 0.036 & 0.258 & 0.043 & 0.195\\
$\gamma_{0}$ & 8.521 & 15.430 & 10.972 & 11.923 & 12.729 & 12.840 & 12.831 & 12.893\\
$\gamma_{1}$ & -0.039 & 1.684 & -0.184 & 0.517 & -0.184 & 0.252 & -0.176 & 0.222\\
$\gamma_{2}$ & -0.016 & 0.977 & -0.100 & 0.348 & -0.090 & 0.154 & -0.079 & 0.127\\
$\gamma_{3}$ & -0.037 & 0.781 & -0.096 & 0.303 & -0.098 & 0.158 & -0.084 & 0.133\\
$\theta$ & -159519.119 & 1344515.437 & -106885.344 & 627588.561 & -69229.439 & 166381.207 & -62599.755 & 74618.949\\
Runs converged (\%) &  & 99.800 &  & 99.500 &  & 97.400 &  & 93.000\\
Sample size ($T$) &  & 100 &  & 200 &  & 1000 &  & 1500\\
\bottomrule
\end{tabular}

  \end{adjustbox}

  \vspace{1em}

  \text{(c) $\sigma = 2.0$}\\[0.5em]
  \begin{adjustbox}{width=0.95\textwidth}
    
\begin{tabular}[t]{lrrrrrrrr}
\toprule
  & Bias & RMSE & Bias & RMSE & Bias & RMSE & Bias & RMSE\\
\midrule
$\alpha_{0}$ & -2.128 & 11.259 & -1.140 & 8.691 & 0.744 & 5.721 & 0.778 & 4.198\\
$\alpha_{1}$ & -0.325 & 1.767 & -0.175 & 1.309 & 0.147 & 1.100 & 0.142 & 0.709\\
$\alpha_{2}$ & -0.019 & 0.139 & -0.007 & 0.145 & 0.011 & 0.076 & 0.010 & 0.066\\
$\alpha_{3}$ & -0.161 & 0.971 & -0.112 & 0.866 & 0.040 & 0.470 & 0.057 & 0.358\\
$\gamma_{0}$ & 9.831 & 18.315 & 10.885 & 13.580 & 12.904 & 13.202 & 12.878 & 13.060\\
$\gamma_{1}$ & -0.065 & 2.134 & -0.125 & 1.066 & -0.207 & 0.404 & -0.179 & 0.330\\
$\gamma_{2}$ & -0.100 & 1.292 & -0.082 & 0.723 & -0.090 & 0.275 & -0.095 & 0.219\\
$\gamma_{3}$ & -0.050 & 1.475 & -0.051 & 0.759 & -0.099 & 0.281 & -0.088 & 0.229\\
$\theta$ & -696456.922 & 5688252.310 & -418083.860 & 2991591.323 & -84373.063 & 140028.575 & -80810.876 & 148172.098\\
Runs converged (\%) &  & 99.100 &  & 99.100 &  & 95.500 &  & 90.600\\
Sample size ($T$) &  & 100 &  & 200 &  & 1000 &  & 1500\\
\bottomrule
\end{tabular}

  \end{adjustbox}
    
  \footnotesize
\end{table}

\begin{table}[!htbp]
  \centering
  \caption{N2SLS with Constraints \eqref{eq:conduct_constraint} for the loglinear model}
  \label{tb:loglinear_loglinear_sigma_0.5_simultaneous_theta_constraint_non_constraint_bias_rmse} 

  \text{(a) $\sigma = 0.5$}\\[0.5em]
  \begin{adjustbox}{width=0.95\textwidth}
    
\begin{tabular}[t]{lrrrrrrrr}
\toprule
  & Bias & RMSE & Bias & RMSE & Bias & RMSE & Bias & RMSE\\
\midrule
$\alpha_{0}$ & -0.087 & 3.688 & 0.145 & 3.117 & 0.027 & 0.975 & -0.066 & 0.764\\
$\alpha_{1}$ & -0.014 & 0.565 & 0.021 & 0.477 & 0.004 & 0.150 & -0.010 & 0.117\\
$\alpha_{2}$ & 0.001 & 0.045 & 0.003 & 0.036 & 0.000 & 0.012 & -0.001 & 0.010\\
$\alpha_{3}$ & -0.004 & 0.317 & 0.009 & 0.263 & 0.002 & 0.090 & -0.004 & 0.073\\
$\gamma_{0}$ & -0.679 & 3.764 & -0.257 & 2.192 & 0.015 & 0.984 & 0.040 & 0.815\\
$\gamma_{1}$ & 0.097 & 0.465 & 0.041 & 0.253 & 0.004 & 0.104 & 0.000 & 0.084\\
$\gamma_{2}$ & 0.049 & 0.290 & 0.023 & 0.170 & 0.002 & 0.073 & 0.003 & 0.060\\
$\gamma_{3}$ & 0.050 & 0.288 & 0.021 & 0.178 & 0.001 & 0.077 & 0.001 & 0.058\\
$\theta$ & -0.063 & 0.389 & -0.068 & 0.333 & -0.043 & 0.199 & -0.032 & 0.167\\
Runs converged (\%) &  & 99.500 &  & 99.700 &  & 100.000 &  & 100.000\\
Sample size ($T$) &  & 100 &  & 200 &  & 1000 &  & 1500\\
\bottomrule
\end{tabular}

  \end{adjustbox}

  \vspace{1em}

  \text{(b) $\sigma = 1.0$}\\[0.5em]
  \begin{adjustbox}{width=0.95\textwidth}
    
\begin{tabular}[t]{lrrrrrrrr}
\toprule
  & Bias & RMSE & Bias & RMSE & Bias & RMSE & Bias & RMSE\\
\midrule
$\alpha_{0}$ & -1.922 & 8.603 & -0.068 & 5.116 & 0.037 & 2.035 & 0.000 & 1.556\\
$\alpha_{1}$ & -0.299 & 1.314 & -0.010 & 0.785 & 0.005 & 0.312 & 0.000 & 0.240\\
$\alpha_{2}$ & -0.013 & 0.104 & -0.002 & 0.063 & 0.001 & 0.024 & 0.000 & 0.019\\
$\alpha_{3}$ & -0.165 & 0.774 & -0.007 & 0.472 & -0.004 & 0.185 & -0.001 & 0.152\\
$\gamma_{0}$ & -1.767 & 14.394 & -1.001 & 6.530 & -0.208 & 1.993 & -0.156 & 1.566\\
$\gamma_{1}$ & 0.255 & 1.949 & 0.132 & 0.838 & 0.034 & 0.229 & 0.027 & 0.174\\
$\gamma_{2}$ & 0.125 & 1.097 & 0.053 & 0.475 & 0.017 & 0.150 & 0.019 & 0.119\\
$\gamma_{3}$ & 0.099 & 0.903 & 0.062 & 0.481 & 0.007 & 0.149 & 0.014 & 0.120\\
$\theta$ & -0.098 & 0.441 & -0.060 & 0.421 & -0.061 & 0.319 & -0.058 & 0.281\\
Runs converged (\%) &  & 98.100 &  & 98.700 &  & 100.000 &  & 100.000\\
Sample size ($T$) &  & 100 &  & 200 &  & 1000 &  & 1500\\
\bottomrule
\end{tabular}

  \end{adjustbox}

  \vspace{1em}

  \text{(c) $\sigma = 2.0$}\\[0.5em]
  \begin{adjustbox}{width=0.95\textwidth}
    
\begin{tabular}[t]{lrrrrrrrr}
\toprule
  & Bias & RMSE & Bias & RMSE & Bias & RMSE & Bias & RMSE\\
\midrule
$\alpha_{0}$ & -4.309 & 16.220 & -2.888 & 12.102 & -0.143 & 5.650 & 0.085 & 3.693\\
$\alpha_{1}$ & -0.657 & 2.489 & -0.448 & 1.840 & -0.024 & 0.874 & 0.012 & 0.568\\
$\alpha_{2}$ & -0.035 & 0.178 & -0.023 & 0.170 & 0.001 & 0.062 & 0.002 & 0.042\\
$\alpha_{3}$ & -0.280 & 1.174 & -0.250 & 1.181 & -0.027 & 0.545 & 0.003 & 0.323\\
$\gamma_{0}$ & 2.206 & 15.660 & -0.576 & 13.314 & -0.897 & 4.882 & -0.646 & 3.317\\
$\gamma_{1}$ & -0.255 & 2.096 & 0.100 & 1.753 & 0.119 & 0.607 & 0.086 & 0.399\\
$\gamma_{2}$ & -0.163 & 1.316 & 0.045 & 1.113 & 0.065 & 0.363 & 0.034 & 0.257\\
$\gamma_{3}$ & -0.155 & 1.669 & 0.068 & 1.127 & 0.061 & 0.386 & 0.041 & 0.271\\
$\theta$ & -0.201 & 0.474 & -0.142 & 0.463 & -0.053 & 0.400 & -0.050 & 0.374\\
Runs converged (\%) &  & 98.100 &  & 97.200 &  & 99.200 &  & 99.400\\
Sample size ($T$) &  & 100 &  & 200 &  & 1000 &  & 1500\\
\bottomrule
\end{tabular}

  \end{adjustbox}
  \footnotesize
\end{table}


\newpage

\newpage

\end{document}